\documentclass[aps,prl,twocolumn,showpacs]{revtex4}
\usepackage{graphicx}
\newcommand{\beq}{\begin{eqnarray}}
\newcommand{\eeq}{\end{eqnarray}}
\begin{document}

\preprint{1}
\title{Memory in the Photon Statistics of Multilevel Quantum Systems}

\author {Felipe Caycedo-Soler and Ferney J. Rodriguez}
\affiliation{Departamento de Fisica, Universidad de Los Andes, A.A. 4976 Bogota,
D.C.,Colombia (South-America)}

\author{Gert Zumofen}
\affiliation{Laboratory of Physical Chemistry, ETH-Zurich, CH-8093 Zurich, Switzerland}

\date{\today}

\begin{abstract}

The statistics of photons emitted by single multilevel systems is
investigated with emphasis on the nonrenewal characteristics of
the photon-arrival times. We consider the correlation between
consecutive interphoton times and present closed form expressions
for the corresponding multiple moment analysis. Based on the
moments a memory measure is proposed which provides an easy way of
gaging the non-renewal statistics. Monte-Carlo simulations demonstrate
that the experimental verification of non-renewal statistics is feasible.

\end{abstract}


\maketitle

\noindent

The arrival times of photons emitted by
single quantum systems have become a task of routine measurements
\cite{KDM,BMO,FSZ,DW,ERA,GSD,ZFB,HSZ,GHH}. Several methods are currently in
use for the analysis of the recorded photon time traces, such as the second order
field-correlation function \cite{KDM,BMO,FSZ,PK,Osa},
photon-number statistics  \cite{Osa1,Ma},
exclusive and nonexclusive
interphoton probability density functions (PDF) \cite{FSZ,Ag1,ZMW,CSV},
or Mandel's Q-function \cite{FSZ,DW,Ma,HB}. Usually, the PDF of photon-arrival
times is considered to depend solely on the arrival time of the previous
photon, assuming tacitly that photon emission is a renewal (semi-Markovian)
process \cite{Ag1}. Consequently, multiple interphoton time PDFs are
factorized and cast into products of one-interphoton time PDFs \cite{FSZ,Ag1}.
In contrast, a nonrenewal process indicates a memory, since the photon-arrival
time PDF depends not only on the arrival time of the previous but also on
the arrival time of the photon before last, and consequently on the
particular realization of the previous photons' time trace \cite{ZMW,CSV,He}.

For ensembles of microscopic photon sources the photon statistics is expected
to be renewal, however, experiments reported on single and coupled
quantum dots \cite{GSD,UML,PAZ,BHH}, on single pairs of coupled molecules
\cite{HSZ,BDM,HZR}, and on two-state dynamics of single molecules
\cite{ZFB} could be considered for the investigation of nonrenewal properties
in the photon statistics. Recently, a renewal indicator was introduced
for the study of conformational fluctuations of single molecules  \cite{Cao}.
This indicator is closely related to Mandel's Q-function and relies on the statistics
of the number of photons recorded in a given time interval. In this paper we consider
another technique which is based on the correlation between
consecutive interphoton times. We apply a multiple-moment analysis
of consecutive interphoton times
and propose a measure ${\cal M}$ for deviations from renewal statistics.

The time evolution of a multilevel quantum system interacting with the
radiation vacuum can be given in terms of the reduced density matrix
$\rho$ by  ($\hbar=1$) \cite{PK,ZMW}
\beq
 \dot \rho = {\cal L}\rho=i[\rho,H]+\sum_{i,j} \gamma_{ij}
 \left( S_i^- \rho S_j^+ - \mbox{ $  \frac 12 $}  \left[S_i^+S_j^- , \rho \right]_+  \right),
 \label{eq1}
\eeq
where the Liouvillian ${\cal L}$ consists of the Hamiltonian $H$, which
includes the interaction with the classical driving field, and of dissipation in
the Lindblad form. As usual, $S_i^-$ ($S_i^+$) are lowering
(rising) operators for the $i$-th transition. $\gamma_{ij}$
denote for $i=j$ the spontaneous emission rates and for $i\ne j$
 cooperative decay rates deviating from zero if the difference
 of the two involved transition frequencies is smaller than
 the inverse radiation-bath correlation time,
 $|\omega_i -\omega_j| \le 1/\tau_{\rm c}$  \cite{He}.
 Assuming the rotating wave approximation, ${\cal L}$ does not depend on
 time so that the evolution of $\rho$ is given by
 $\rho(t) = e^{{\cal L}t}\rho(0)$, where $\rho(0)$ denotes the state at time zero.

For the description of state collapses upon photon detection,
several theoretical approaches, pioneered by the Monte-Carlo wave
function technique \cite{DCM}, were developed. These approaches
rely on quasi-continuous photon-emission measurements to introduce
system states conditioned on whether a photon is detected or
not \cite{He,BH}. Accordingly, the Liouvillian  is split into two
terms \cite{He}
 \beq
   {\cal L} = {\cal L}_{\rm c} + {\cal R} ~,
 \label{eq2}
\eeq
where ${\cal L}_{\rm c}$ governs the
time evolution of the conditioned and non-normalized density
matrix $\rho_{\rm c}(t) = e^{{\cal L}_{\rm c} t} \rho(0) = {\cal
U}_{\rm c}(t)\rho(0)$, subject to a zero-photon outcome of the
measurement up to time $t$. The second term in Eq. (\ref{eq2}) represents the
collapse (reset, recycling) operator ${\cal R}$ \cite{He}  to reset the density
matrix upon a photon detection event \cite{CSV,SM}
\beq
  {\cal R} \rho = \eta  {\sum}_{i, j} \gamma_{ij}  S_i^- \rho S_j^+ ~.
 \label{eq3}
\eeq
The dimensionless detection efficiency $\eta$ is introduced
to account for the fact that a state collapse
takes exclusively place when the emitted photon is also detected \cite{CSV}.
According to Eq. (\ref{eq2}), also the conditioned Liouvillian
${\cal L}_{\rm c}= {\cal L -R}$ depends on $\eta$ in a unique way.
${\cal R}$ operating on $\rho_{\rm c}(t)$ at random times generates a
stochastic process and thus the average survival probability $P_0(t)$
of no-photon detection up to time $t$ is \cite{PK}
\beq
 P_0 (t) &=& {\rm Tr} \left\{  \rho_{\rm c}(t) \right\}
 = {\rm Tr} \left\{ {\cal U}_{\rm c}(t) \rho_0 \right\} ,
 \label{eq4}
\eeq
provided that a photon was recorded at time zero. Correspondingly, $\rho_0$
is the average state just after photon detection and is given by normalizing
the collapsed stationary state,
$\rho_0= {\cal R} \rho^{\rm ss} / {\rm Tr} \left\{{\cal R} \rho^{\rm ss} \right\}$,
where the stationary state satisfies ${\cal L} \rho^{\rm ss} = 0$.

\begin{figure}
\includegraphics[width=0.8\columnwidth]{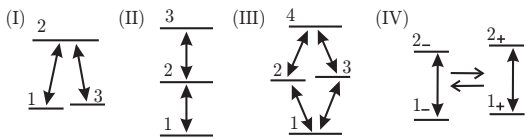}
\caption{ Multi-level systems under consideration: (I) three-level $\Lambda$-system,
(II) cascade three-level system, (III) four-level system motivated by a pair of
interacting two-level systems, and (IV) a two level system jumping stochastically
and radiationless between two states. Heavy arrows for laser-light driven and
spontaneous radiation transitions and light arrows for radiationless transitions.}
\end{figure}

Recording the times of state collapses generated by repeated application of the
operator ${\cal RU}_{\rm c}(t)$  mimics the time traces of photon detection
in a particular single quantum system experiment. The conditional density
matrix right after the  $n$-th photon detection of a sequence of exclusive
detection times $\{t_1,\cdots,t_n\}$ with $t_i\ge t_{i-1}$
is then given by
\beq
  \rho_{\rm c}(t_1,t_2,.., t_{n})
  = \left[{\cal T}_+ {\prod}_{i=1}^n  {\cal R} {\cal U}_{\rm c}(t_{i}-t_{i-1} ) \right]  \rho_0 ~,
 \label{eq5}
\eeq
where the time ordering operator ${\cal T}_+ $ ensures that the operator at
the latest time is on the far left. The trace of
$\rho_{\rm c}(t_1,t_2,.., t_{n})$ in Eq. (\ref{eq5})
provides the detection PDF of a particular time sequence
\beq
    p_n(\tau_1,\tau_2,\cdots,\tau_{n})=
    \mbox{Tr} \left\{\left[{\cal T}_+  {\prod}_{i=1}^n {\cal R}{\cal U}_{\rm c}(\tau_i) \right]
     \rho_0\right\}~,
 \label{eq6}
\eeq
where $\tau_i= t_i-t_{i-1}$ are interphoton times. Furthermore, the PDF $P_2(t)$
of detecting a second photon at time $t$, given a detection event at any previous
instance, follows from summing up all possible realizations of two consecutive
interphoton times \cite{ZMW}
\beq
 P_2(t)=\int_0^{t}   p_2(t-\tau_1,\tau_1) \, {\rm d} \tau_1
 \label{eq7}~.
\eeq
Generally, referring to Eq. (\ref{eq6}) the PDF $P_n(t)$ of the $n$-th photon at time $t$
results from the $n-1$ fold convolution of the operator ${\cal
R}{\cal U}_{\rm c}(t)$, where for completeness, $P_1(t) = p_1(t)$. For the
quantitative analysis of $p_n$, we examine the moments to order $m_i,i=1,\cdots,n$
for $n$
consecutive detection intervals. These moments can readily be calculated using
a moment generating function technique in several dimensions
\beq
 \lefteqn{\mu_{m_1,..,m_n} = \left( {\prod}_{i=1}^n \int_0^{\infty}  {\rm d} \tau_i  \,\tau_i^{m_i} \right)
      p_n(\tau_1,\cdots,\tau_{n}) }   \nonumber \\
 =  && \mbox{Tr}\left\{ \left( {\cal T}_+{\prod }_{i=1}^n (-1)^{(m_i+1) } m_i !{\cal R}{\cal L}_{\rm c}^{-(m_i+1)}
      \right)   \rho_0 \right\} ,
 \label{eq9}
\eeq
where, recalling, the ordering operator ${\cal T}_+ $ ensures that the operator
at the latest time is on the far left. Eq. (\ref{eq9}) allows for an easy
numerical calculation of multiple moments. In case of renewal,
${\cal R} \rho(t)$ does not depend on $t$, in other words the state after a
collapse is independent of the state just before the collapse. Consequently,
$p_n(\tau_1,\cdots\tau_{n})$ of Eq. (\ref{eq6})  can be factorized in terms
of the one-interphoton time PDF,
$p_n^{\rm R}(\tau_1,\cdots,\tau_{n}) = \prod_{i=1}^n p_1( \tau_i)$, where
the superscript R denotes renewal. Furthermore, the arrival PDF of the
second photon is $P_2^{\rm R}(t) = p_1(t)*p_1(t)$, where $*$ indicates
convolution and  multiple moments reduce to products of individual moments
 \beq
 \mu^{\rm R}_{m_1,\cdots,m_n}= {\prod}_{i=1}^n \langle \tau^{m_i} \rangle
    = {\prod}_{i=1}^n \mu_{m_i} .
 \label{eq10}
\eeq

Differences between the PDFs  $p_n(\tau_1,\cdots,\tau_{n})$ and $p_n^{\rm
R}(\tau_1,\cdots,\tau_{n})$, $P_2(t)$ and $P_2^{\rm R}(t)$, or between the moments
of Eqs. (\ref{eq9}) and (\ref{eq10}) may be used to demonstrate whether  the
initial state $\rho_0$ is recovered after photon emission and the process is
renewal or whether the state-resetting depends on the current state and the
process is nonrenewal. The deviation from renewal is a signature of the lack of
information about the system state after photon emission and indicates the memory
present in the correlation between consecutive photon-arrival times. Envisaging
the experimental verification of NRS we concentrate on two consecutive time
intervals and propose the following measure
\beq
 {\cal M} =  \mu_{1,1} / \mu_{1}^2 -1 ~,
 \label{eq11}
\eeq
which can be determined directly from the experimental time traces and can easily be predicted
using Eq. (\ref{eq9}). $\cal M$ takes on both signs and an analysis of bi-valued
waiting-time sequences indicate that tentatively $\cal M$ is negative
when shorter and longer waiting times are likely to
occur alternatingly and is positive when both, shorter and longer waiting times
are likely to be bunched.

The multiple moments of Eq. (\ref{eq9}) and the measure $\cal M$ of Eq. (\ref{eq11})
represent the main result of this paper and we consider ${\cal M}$ as a characteristic
quantity complementary to other statistical measures, e.g. the photon coincidence
probability (PCP) $g^{(2)}(0)$. We also studied the covariance of the functions
$P_2(t)$ and $P_2^{\rm R}(t)$, however, such an analysis requires binning of the
time traces and is therefore less direct for gaining information about the memory
in the photon statistics.

\begin{figure}
\includegraphics[width=1.\columnwidth]{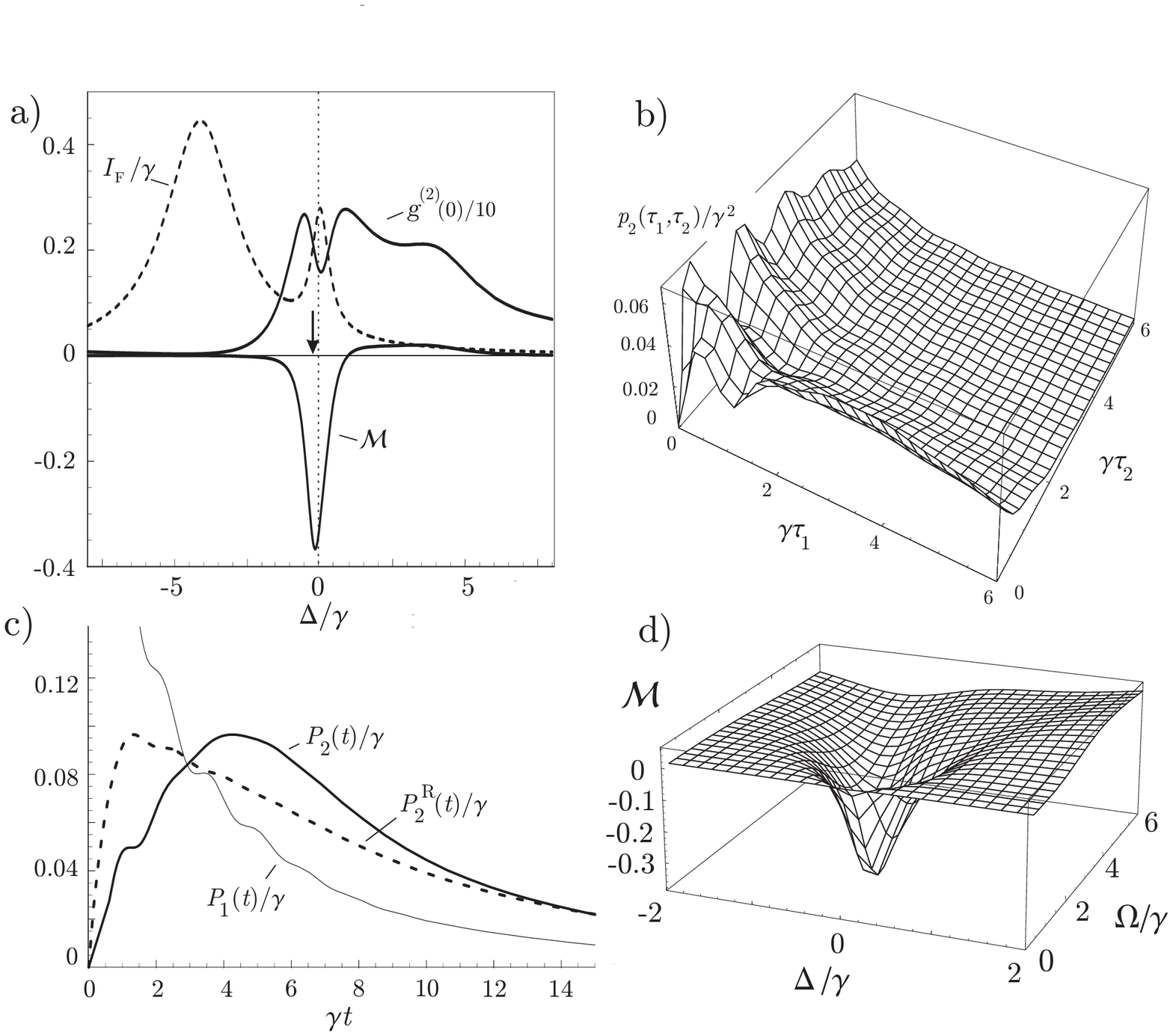}
\caption{  The three-level cascade system. a) Memory ${\cal M}$, PCP $g^{(2)}(0)$,
and fluorescence-excitation intensity $I_{\rm F}$ as a function of the
laser detuning $\Delta$. The arrow indicates the $\Delta$ value used in b) and c).
b) Renewal and nonrenewal second-photon arrival PDF $P_2^{\rm R}(t)$ and $P_2(t)$,
respectively, and one-photon arrival PDF $P_1(t)$. c) Two-photon arrival
PDF $p_2(\tau_1,\tau_2)$. d) Memory $\cal M$ as a function of the detuning
$\Delta$ and the Rabi frequency ${\mathit \Omega}$.}
\end{figure}

To illustrate the NRS in photon counting we consider a representative set of level
schemes, shown in Fig. 1. For the three-level $\Lambda$-system (I) renewal applies,
because once a photon is detected the system is reset to the same mixed state, no
matter the state before emission. In this respect, the $\Lambda$-system equals
a two-level system which always collapses to the ground state. In contrast, NRS
arises for systems (II-IV): For the three level cascade (II), upon the detection
of a spectrally unresolved photon, the populations and coherences of the states
$|1\rangle$ and $|2\rangle$ become proportional to the ones of  $|2\rangle$ and
$|3\rangle$ prior to emission, respectively \cite{He}. In the four level
cascade (III), the reset operator ladders populations and coherences down
the levels: $|4\rangle\rightarrow(|3\rangle,|2\rangle )\rightarrow |1\rangle$.
The non-cascade system (IV) shows a TLS flipping stochastically between two states.
The flipping may be associated with changes of spectral and dynamical properties
so that bunching of short and long interphoton times and thus NRS results.
Summarizing, if the state after emission depends on the state prior to emission, the
photon time traces obey NRS. The PCP can be discussed accordingly, namely, for system
(I) it is zero, and is non-zero for systems (II) and (III). However, for system (IV)
the PCP is zero although the process is nonrenewal in general.

\begin{figure}
\includegraphics[width=1.\columnwidth]{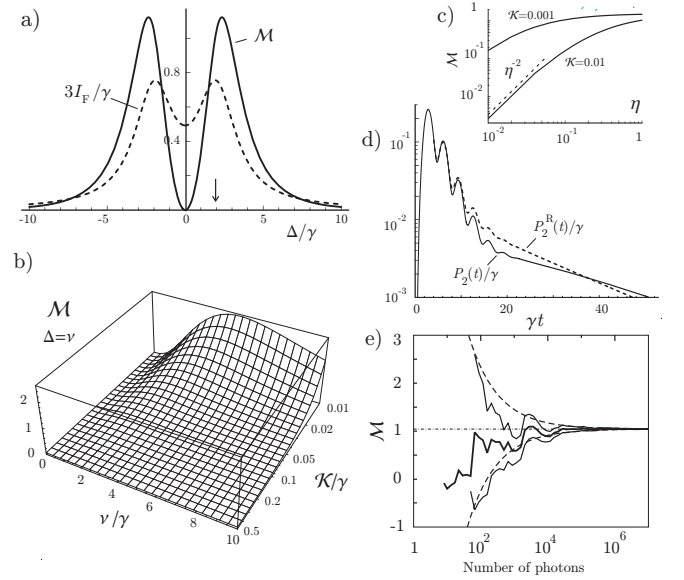}
\caption{ Jumping two-level system. a) $I_{\rm F}$ scaled as indicated and $\cal M$
as a function of the laser detuning, similarly as in Fig. 2a.
The arrow indicates the detuning used in Fig. 2d. b) $\cal M$ as
function of $\nu$ and $\cal K$ on linear and log scales, respectively.
c) $\cal M$ as a function of the detection efficiency $\eta$
on log-log scales for two values of $\cal K$. The dashed line for the $\eta^{-2}$
dependence. d) Second-photon arrival time PDFs $P_2^{\rm R}(t)$ and $P_2(t)$.
e) $\cal M$ and confidence intervals ${\cal M} \pm \sigma_{\cal M}$ as a function
of the number of detected photons. Wiggly lines for Monte-Carlo simulations and
dashed lines for predictions. The dash-dotted line gives the predicted value
${\cal M} = 1.04 $. $\Delta=\nu$ in panels b)-e). }
\end{figure}

We next discuss the level schemes (II-IV) in more detail. The Hamiltonian in Eq.
(\ref{eq1})  can be written as $H = H_0  +\sum_i \frac 12 {\Omega_i} (S_i^+ + S_i)$,
where $\Omega_i$ is the Rabi frequency of the $i$-th transition resulting from the
interaction with the classical driving field. For level scheme (II) we write for the
zero-order Hamiltonian
 \beq
  H_0^{\rm (II)} &=  -  \delta_1 S_1^+  S_1^- - (\delta_1 + \delta_2)  S_2^+ S_2^-,
\label{eq12}
\eeq
where $S_1^- = |1\rangle \langle 2|$ and  $S_2^- = |2\rangle\langle 3|$ are lowering
operators with rising operators defined accordingly. $\delta_1 = \omega_{\rm L}-\omega_{21}$
and $\delta_2 = \omega_{\rm L}-\omega_{32}$
are differences between the laser frequency $\omega_{\rm L}$ and transition
frequencies $\omega_{ij}$. Results are shown in Fig. 2 for the parameters:
${\mathit \Omega}_1={\mathit \Omega}_2={\mathit \Omega}=2\gamma,
\gamma_{11} = \gamma_{22} = \gamma, \gamma_{12} = \gamma_{21} =0,
\omega_{32}-\omega_{21} = 8 \gamma$, and $\eta =1$.
In Fig. 2a the excitation fluorescence intensity
$I_{\rm F} = {\rm Tr} \left\{ {\cal R }\rho^{\rm ss} \right\}$, the PCP
$g^{(2)}(0) =  {\rm Tr} \left \{ {\cal R}^2 \rho^{\rm ss} \right\}/
I_{\rm F}^2 $, and ${\cal M}$ are compared as functions of the detuning
$\Delta = \frac 12 (\delta_1+\delta_2)$. The fluorescence shows a maximum
located approximately at resonance with the the lower transition
($\delta_1 \simeq 0$), followed by a peak at $\Delta \simeq 0$,
where coherent two-photon absorption and cascade emission are likely to occur.
The intensity at resonance with the upper transition ($\delta_2 \simeq 0$)
is weak because of weak pumping of level 2. The PCP indicates photon
antibunching in the range of the lower transition and photon bunching
in the range of two-photon absorption and of the upper transition. In
agreement with the above discussion, $\cal M$ takes on negative values
in the range of $\Delta \simeq 0$, where alternating short and long
interphoton times are probable. At resonance with the upper transition,
$\cal M$ is  weakly positive indicating minor bunching of short and long
waiting times. $P_2(t)$ and $P_2^{\rm R}(t)$ displayed in Fig. 2b deviate
considerably from each other and similarly a strong asymmetry is apparent
in the PDF $p_2(\tau_1,\tau_2)$ upon interchanging $\tau_1$ and $\tau_2$
demonstrated in Fig. 2c. In Fig. 2d $\cal M$ is monitored as a function
of $\Delta$ and ${\mathit \Omega}$. A minimum close to $\Delta\simeq0$
and ${\mathit \Omega} \simeq 2 \gamma$ is clearly visible. We have found
that $|\cal M|$ drops roughly as $\eta^{-2}$ so that the experimental
verification of NRS requires a high detection efficiency.

Motivated by recent investigations of pairs of identical and interacting
quantum systems \cite{HSZ,BHH,HZR} we studied system (III). Depending on
the parameters the results (not shown here) were similar to those of
system (II) which is obvious when states 2 and 3 are superposition Dicke
states \cite{BH,AFS}, so that the behavior is governed by the 1$\leftrightarrow$2
and 2$\leftrightarrow$4 transitions or 1$\leftrightarrow$3 and 3$\leftrightarrow$4
transitions.

We finally report on the non-cascade, jumping two-level system (IV) where the
$\pm$ states indicate for instance two different molecular, or lattice nuclear
configurations \cite{ZFB, HB}, spectral diffusion of ultracold molecules in condensed media
\cite{Osa2}, or two different spin configurations. The dynamics is described by extending the
density operator, $\rho=(\rho_-,\rho_+)^{\rm T}$, and correspondingly the Liouvillian
and reset operators \cite{HB}
\beq
  {\cal L}^{\rm (IV)} = \left( \begin{array}{cc} {\cal L}_{ -}-{\cal K}_-
   & {\cal K}_+ \\ {\cal K}_- &{\cal L}_{+}-{\cal K}_+ \end{array} \right) ,
   {\cal R} = \left( \begin{array}{cc} {\cal R}_-
   & 0 \\ 0 &{\cal R}_+ \end{array} \right),~~
\eeq
where ${\cal K}_\pm$ denote the jumping rates between the two states and where
${\cal L}_\pm$ and ${\cal R}_\pm$ are the Liouvillian and resetting operators
of the two states. For illustration
we assume the full symmetric case where only the transition frequencies are
different for the two sates. Thus the system is described by the Hamiltonian
$H_\pm = (-\Delta \pm \nu) S^+_\pm \ S_\pm + \frac 12 {\mathit \Omega} (S_\pm + \ S^+_\pm)$,
where $\Delta$ and $\nu$ are laser detuning and frequency displacements from
the transition center, respectively. Furthermore, $\gamma_\pm = \gamma$,
${\cal R}_\pm = {\cal R}$, and ${\cal K}_\pm = {\cal K}$.
Numerical results are shown in Fig. 3 for the parameters
${\mathit \Omega}=2 \gamma, \nu= 2 \gamma, \eta = 1, {\cal K} = \gamma/100$,
except when they appear as variables or are specially indicated.
${\cal M}$ is positive throughout all calculations and peaks close to the
resonances of the two states. Fig. 3b indicates large positive $\cal M$ for
${\cal K} \ll \gamma$ and for $|\Delta| \simeq |\nu|$. Fig. 3c shows how
the $\eta$ dependence of $\cal M$ crosses over to the asymptotic
$\eta^{-2}$ behavior and how the crossover is shifted to lower values of
$\eta$ with decreasing $\cal K$. The second-photon arrival PDFs
$P_2^{\rm R}(t)$ and $P_2(t)$ in Fig. 3d differ only at longer times which
indicates that these quantities are not appropriate for providing evidence
of NRS.

To demonstrate the experimental feasibility of measuring $\cal M$, we report
Monte-Carlo simulation results \cite{PK,DCM}. Choosing a random number $r$
distributed uniformly in $[0,1]$, the detection time $t_n$ of the n-th photon
follows from  the condition
\beq
 {\rm Tr} \left\{ {\cal U}_{\rm c}(t_n-t_{n-1})
      \hat \rho_{\rm c}(t_1,\cdots,t_{n-1}) \right\} = r ~,
 \label{eq13}
\eeq
where $\hat \rho_{\rm c}$ is the normalized conditioned density matrix of Eq. (5).
By resetting and normalizing the state at $t_n$, the initial state of the next
interphoton cycle is obtained. Fig. 3e shows, how $\cal M$ converges as a
function of the photon number to the predicted value. Also presented are
confidence intervals $\cal M\pm \sigma_{\cal M}$, which are estimated from
the variance, assuming statistical independence of the moments:
$\sigma^2_{\cal M} =   \sigma^2_{\mu_{1,1}} / \mu_{1}^4 + 4 (  \mu_{1,1} / \mu_1^3  )^2
\sigma^2_{\mu_{1}}$, where $\sigma^2_{\mu_{1}} = \mu_{2} - \mu_{1}^2$ and
$\sigma^2_{\mu_{1,1}} = \mu_{2,2} - \mu_{1,1}^2$.

The experimental investigation of the NRS requires the measurement of two
consecutive intervals, so that the arrival times of three consecutive photons
have to be recorded. Depending on the time scale, this can be achieved using
a single detector, however, for time scales shorter than the detectors' dead
time, at least three detectors are needed. A comprehensive description of
experimental data has to account for the detectors' dead time and for the ubiquitous
background photons.

In conclusion, we have shown that NRS is plausible in the fluorescence of
multi-level systems and that the indicator $\cal M$, proposed for the identification
of the nonrenewal property, is experimentally feasible. For cascade systems
$\cal M$ may be small at low detection efficiency, so that advanced experimental
techniques are required, while for non-cascade multi-level systems $\cal M$ may
be large also at low detection efficiency.

We thank  V. Sandoghdar  for stimulating discussions. F.C.S. thanks the ETH-Zurich
for the hospitality and the financial support from COLCIENCIAS No. 1204-05-11408
and Banco de la Rep\'ublica.


\begin{thebibliography}{18}
\bibitem{KDM} H.J. Kimble, M. Dagenais, L. Mandel, Phys. Rev. Lett. {\bf 39}, 691 (1977).
\bibitem{BMO} Th. Basch\'e et al.,  Phys. Rev. Lett. {\bf 69} 1516  (1992).
\bibitem{FSZ} L. Fleury et al., Phys. Rev. Lett. {\bf 84},  1148 (2000).
\bibitem{DW} F. Diedrich and H. Walther,  Phys. Rev. Lett. {\bf 58} 203 (1987).
\bibitem{ERA} J. Enderlein, D. I. Robbins, W. P. Ambrose, and R. A. Keller,
J. Phys. Chem. A 102, 6089 (1998).
\bibitem{GSD} B.D. Gerardot et al., Phys. Rev. Lett. {\bf 95},  137403 (2005).
\bibitem{ZFB} A. Zumbusch et al., Phys. Rev. Lett. {\bf 70}, 3584 (1993).
\bibitem{HSZ} C. Hettich et al., Science {\bf 298}, 385 (2002).
\bibitem{GHH} G. Zumofen, J. Hohlbein, C.G. H\"ubner, Phys. Rev. Lett. {\bf 93}, 260601 (2004).
\bibitem{Osa} I.S. Osad'ko, J. Luminisc. {\bf 87}, 184 (2000).
\bibitem{PK} M.B. Plenio and P.L. Knight, Rev. Mod. Phys. {\bf 70}, 101 (1998).
\bibitem{Osa1} I.S. Osad'ko, JETP Lett. {85}, 550 (2007).
\bibitem{Ma} L. Mandel, Opt. Lett. {\bf 4}, 205 (1979).
\bibitem{Ag1} G.S. Agarwal, Phys. Rev. A {\bf 15}, 814 (1977).
\bibitem{ZMW} P. Zoller, M. Marte, and D.F. Walls, Phys. Rev. A {\bf 35}, 198 (1987).
\bibitem{CSV} H.J. Charmichael et al.,  Phys. Rev. A {\bf 39} 1200  (1989).
\bibitem{HB} Y. He and E. Barkai,  Phys. Rev. Lett. {\bf 93} 68302 (2004).
\bibitem{He} G.C. Hegerfeldt, Phys. Rev. A {\bf 47}, 449 (1993).
\bibitem{UML} T.  Unold, et al., Phys. Rev. Lett. {\bf 94}, 137404 (2005).
\bibitem{PAZ} J. Persson et al., Phys. Rev. B {\bf 69}, 233314 (2004).
\bibitem{BHH} M. Bayer, et al, Science {\bf 291}, 451 (2001).
\bibitem{BDM} A.J. Berglund, A. C. Doherty, and H. Mabuchi, Phys. Rev. Lett. {\bf 89}, 068101 (2002).
\bibitem{HZR} C.G. H\"ubner et al., Phys. Rev. Lett. {\bf 91}, 093903 (2003).
\bibitem{Cao} J. Cao, J. Phys. Chem. B {\bf 110}, 19040 (2006).
\bibitem{DCM} J. Dalibard et al., Phys. Rev. Lett. {\bf 68}, 580 (1992).
\bibitem{BH} A. Beige and G.C. Hegerfeldt, Phys. Rev. A {\bf 58}, 4133 (1998).
\bibitem{SM} F. Sanda and S. Mukamel, Phys. Rev. A {\bf 71}, 33807 (2005).
\bibitem{AFS} U. Akram, Z. Ficek, and S. Swain,  Phys. Rev. A {\bf 62}, 13413 (2000).
\bibitem{Osa2} I.S. Osad'ko and E. V. Kohts, Optics and Spectroscopy {\bf 94}, 949 (2003).

\end{thebibliography}
\end{document}